\documentclass[aps,prl,twocolumn,superscriptaddress,preprintnumbers,%
               showpacs,nofootinbib,floatfix]{revtex4}
\newcommand{\PRE}[1]{}       

\usepackage{bm}
\usepackage{epsfig}
\usepackage{hyperref}

\newcommand{\postscript}[2]{\setlength{\epsfxsize}{#2\hsize}
   \centerline{\epsfbox{#1}}}

\def\to{\rightarrow}

\def\bi{\begin{itemize}}
\def\ei{\end{itemize}}

\def\ta{\tilde a}

\def\tb{\tilde b}

\def\tst{\tilde t}

\def\tg{\tilde g}

\def\tq{\tilde q}
\def\tw{\widetilde W}
\def\tz{\widetilde Z}

\def\alt{\lesssim}
\def\agt{\gtrsim}
\def\be{\begin{equation}}
\def\ee{\end{equation}}
\def\bea{\begin{eqnarray}}
\def\eea{\end{eqnarray}}

\newcommand\sjp[3]{{\it Sov.\ J.\ Nucl.\ }{\bf #1} (#2) #3}

\newcommand\plb[3]{{\it Phys.\ Lett.\ }{\bf B#1} (#2) #3}
\newcommand\jhep[3]{JHEP {\bf #1} (#2) #3}
\newcommand\npb[3]{{\it Nucl.\ Phys.\ }{\bf B#1} (#2) #3}

\newcommand\prD[3]{{\it Phys.\ Rev.\ }{\bf D#1} (#2) #3}
\newcommand\prL[3]{{\it Phys.\ Rev.\  Lett.\ }{\bf #1} (#2) #3}

\begin{document}

\preprint{OU-HEP-130815, KIAS-P13048}

\title{
\PRE{\vspace*{1.5in}}
Mainly axion cold dark matter from natural supersymmetry
\PRE{\vspace*{0.3in}}
}
\author{Kyu Jung Bae}
\affiliation{Dept. of Physics and Astronomy,
University of Oklahoma, Norman, OK, 73019, USA
\PRE{\vspace*{.1in}}
}
\author{Howard Baer}
\affiliation{Dept. of Physics and Astronomy,
University of Oklahoma, Norman, OK, 73019, USA
\PRE{\vspace*{.1in}}
}
\author{Eung Jin Chun}
\affiliation{Korea Institute for Advanced Study, Seoul 130-722, Korea
\PRE{\vspace*{.1in}}
}


\begin{abstract}
\PRE{\vspace*{.1in}}
By eschewing fine-tuning from the electroweak and QCD sectors of supersymmetry (natural supersymmetry or SUSY),
and by invoking the Kim-Nilles solution to the SUSY $\mu$ problem,
one is lead to models wherein the dark matter is comprised of a mixture of axions and Higgsino-like WIMPs.
Over a large range of Peccei-Quinn breaking scale $f_a\sim 10^9-10^{12}$ GeV, one then expects
about 90-95\% axion dark matter. In such a scenario, both axion and WIMP direct detection may be expected.

\end{abstract}


\maketitle

The recent discovery of a Higgs-like boson with mass $m_h\simeq 125$ GeV at the
CERN LHC is a triumph of modern particle physics~\cite{lhchiggs}. But it brings with it a conundrum:
why is the Higgs mass so small? In the Standard Model (SM), one may calculate
\be
m_h^2(\text{phys})=m_h^2(\text{tree})+\delta m_h^2
\ee
where the radiative correction $\delta m_h^2\sim -\frac{3f_t^2}{8\pi^2}\Lambda^2+\cdots $
where $f_t\sim 1$ is the top quark Yukawa coupling and $\Lambda$ is a high energy cutoff/regulator
which denotes the limit of validity of the effective theory.
If the SM is to be valid at energy scales $\Lambda$ far beyond
$m_{\text{weak}}\sim 100$ GeV,
then an enormous fine-tuning will be required to maintain $m_h\sim 125$ GeV.

Supersymmetric (SUSY) theories of particle physics provide all-orders cancellations of the quadratic
divergences thus stabilizing the Higgs mass. In SUSY, $\delta m_h^2$ is instead logarithmically divergent,
and includes terms such as
\be
\delta m_h^2(\text{SUSY})\sim -\frac{3f_t^2}{8\pi^2}(m_{Q_3}^2+m_{U_3}^2+A_t^2)\ln\left(\Lambda^2/m_{\text{SUSY}}^2\right)
\label{eq:mh_susy}
\ee
where $m_{Q_3}^2$, $m_{U_3}^2$ and $A_t^2$ are soft SUSY breaking terms related to the top- and 
bottom- squark masses and $m_{\rm SUSY}\sim\sqrt{m_{\tst_1}m_{\tst_2}}$.
Requiring $\delta m_h^2\alt 10 m_h^2$ for $\Lambda$ as high as $m_{\rm GUT}$ 
implies light 3rd generation squarks $m_{\tst_{1,2}},m_{\tb_1}\lesssim 200$ GeV~\cite{kn,ns}, although 
this rather severe fine-tuning measure ignores non-independent contributions to $m_h^2$ which lead to large
cancellations~\cite{comp}.
A model-independent, more conservative measure which allows for cancellations within $m_h^2$ requires that 
the magnitude of all {\it weak scale} contributions to $m_h^2$ (or $m_Z^2$) be comparable to $m_h^2$ (or $m_Z^2$).
For instance, the $Z$ mass is given by
\be
\frac{m_Z^2}{2} \simeq -m_{H_u}^2-\Sigma_u^u-\mu^2
\label{eq:mZsSig}
\ee
which is valid for the ratio of Higgs field vacuum expectation values (vevs) $\tan\beta\equiv v_u/v_d$
over its typical range of $3-60$.
Here, $m_{H_u}^2$ is a soft SUSY breaking Higgs mass,
$\mu$ is the superpotential Higgs/Higgsino mass and
$\Sigma_u^u$ collects various radiative corrections
(expressions are provided in the Appendix of Ref.~\cite{rns}).
For many models, $m_{H_u}^2$ is driven radiatively to large negative values 
$|m_{H_u}^2|\gg m_Z^2$ signaling the breakdown of electroweak symmetry. 
A large positive value of $\mu^2$ must be imposed (fine-tuned)  to obtain the
measured $Z$-mass, $m_Z\simeq 91.2$ GeV. 
(Alternatively, if $|m_{H_u}^2|$ is small but the $\Sigma_u^u$ terms become 
large positive, $\Sigma_u^u\gg m_Z^2$, 
again a large value of $\mu^2$ must be imposed, leading again to fine-tuning.)
To avoid large uncorrelated cancellations (fine-tuning) in the $Z$ mass, 
then one expects $|\mu |$ and $|m_{H_u}|\sim m_Z$, 
or of order $100-200$ GeV~\cite{ltr,rns}.
In addition, requiring the dominant radiative corrections 
\bea
\Sigma_u^u(\tst_{1,2} )&=&\frac{3}{16\pi^2}F(m_{\tst_{1,2}}^2)\times \nonumber \\
\left[ f_t^2-g_Z^2\right. &\mp &
\left.\frac{f_t^2 A_t^2-8g_Z^2(\frac{1}{4}-\frac{2}{3}x_W)\Delta_t}{m_{\tst_2}^2-m_{\tst_1}^2}
\right]
\label{eq:Siguu}
\eea
(where $\Delta_t=(m_{\tst_L}^2-m_{\tst_R}^2)/2+m_Z^2\cos 2\beta(\frac{1}{4}-\frac{2}{3}x_W)$, $g_Z^2=(g^2+g^{\prime 2})/8$, 
$x_W\equiv \sin^2\theta_W$ and where $F(m^2)=m^2(\log \frac{m^2}{Q^2}-1)$ 
with $Q^2\simeq m_{\tst_1}m_{\tst_2}$ the optimized scale choice for
minimization of the scalar potential)
to be $\alt 100-200$ GeV then requires highly mixed top squarks with mass 
$m_{\tst_1}\sim 1-2$ TeV and $m_{\tst_2}\sim 3-4$ TeV~\cite{ltr,rns}.\footnote{Using the full radiative corrections, then 
large values of weak scale $A_t$ suppress $\Sigma_u^u (\tst_1)$ via 
the square bracket in Eq.~(\ref{eq:Siguu}) and via the $F$ function for $\Sigma_u^u(\tst_2 )$.
The suppression due to mixing then allows for much larger stop masses (around the few TeV scale) 
than are found in generic natural SUSY models, where it is often claimed 
that $m_{\tst_{1,2}}$ need be $\alt 200-500$ GeV~\cite{ns}.}
Such highly mixed top squark masses lift $m_h$ 
into the 125 GeV range (even with stops as light as a few TeV) since $m_h$ is maximized for large mixing~\cite{mhiggs}. 
The TeV-scale top squark masses are also heavy enough to suppress anomalous contributions to $b\to s\gamma$ decay 
and to avoid recent LHC null results for top squark searches~\cite{rns}.

The above features have enormous implications for SUSY phenomenology: in this case, the
lightest SUSY particle, often touted as a WIMP dark matter candidate, is expected to be largely
Higgsino-like since the Higgsino mass is $\simeq |\mu |$ and $|\mu |\sim m_Z$. 
Since Higgsino-like WIMPs couple directly to the Goldstone components of the gauge bosons, 
they can annihilate with large rates into $WW$ and $ZZ$ final states. Since the WIMP thermal 
relic abundance is inversely proportional to its annihilation rate, then Higgsinos with mass 
$m_{\tz_1}\sim 100-200$ GeV develop a relic density of $\Omega_{\tz_1}h^2\sim 0.005-0.01$, 
{\it i.e.} typically a factor $10-15$ below the WMAP/Planck measured value~\cite{bbm}.
Supersymmetric models with low fine-tuning (not too heavy 3rd generation squarks and low $\mu$)
are referred to as {\it natural SUSY}~\cite{kn,ns,ltr,rns}
since they are devoid of large electroweak fine-tuning.

A further possibility for fine-tuning occurs in the QCD sector. To implement 't~Hooft's solution
to the $U(1)_A$ problem~\cite{peccei} ({\it i.e.}~why there are three and not four light pions),
the term
\be
\frac{\bar{\theta}}{32\pi^2}F_{A\mu\nu}\tilde{F}_A^{\mu\nu}
\label{eq:FF}
\ee
should occur in the QCD Lagrangian, where $\bar{\theta}=\theta +\text{arg}(\text{det}{\cal M})$,
${\cal M}$ is the quark mass matrix, $F_{A\mu\nu}$ is the gluon field strength and
$\tilde{F}_A^{\mu\nu}$ is its dual.
Measurements of the neutron electric dipole moment (EDM) require $\bar{\theta}\alt 10^{-10}$
so that one might require an enormous cancellation within $\bar{\theta}$~\cite{axreview}. Alternatively,
the PQWW solution~\cite{pqww} introduces an {\it axion} field $a$; the additional axion contributions to
Eq.~(\ref{eq:FF}) allow for $\bar\theta$ to dynamically settle to zero, thus solving
the so-called strong $CP$ problem.

In SUSY theories, the axion enters as one element of an axion {\it superfield} which
necessarily contains also a spin-0 $R$-parity even saxion $s$ and a spin-$1/2$ $R$-parity-odd
axino $\ta$. Calculations of the saxion and axino masses within the context of supergravity~\cite{maxino}
imply $m_s\sim m_{\ta}\sim m_{3/2}$ where the gravitino mass $m_{3/2}$ is expected to be of
order the TeV scale.
If the lightest neutralino ({\it e.g.} the Higgsino $\tz_1$) is the lightest SUSY particle (LSP)
in $R$-parity conserving theories, then one would expect dark matter to be comprised of {\it two particles}:
the axion as well as the Higgsino-like SUSY WIMP.
The saxion and axino couplings to matter are suppressed by the PQ breaking scale
$f_a$ which may range from $f_a\sim 10^9-10^{16}$ GeV~\cite{axreview}.
While the saxion and axino are expected
to play little or no role in terrestrial experiments, they can have an enormous impact on
dark matter production in the early universe.

The PQ symmetry required to solve the strong $CP$ problem can be implemented in two ways.
In the SUSY KSVZ model~\cite{ksvz,kim}, the axion superfield couples to exotic heavy quark/squark superfields
$Q$ and $\bar{Q}$ which carry PQ charges. The loop-induced axino-gluino-gluon coupling
leads to a thermal axino production rate proportional to  the re-heat temperature $T_R$ at the end
of inflation. It also allows for axino decays $\ta\to g\tg$, $\ta\to\gamma\tz_i$ or $\ta\to Z\tz_i$
(with $i=1-4$).
In SUSY KSVZ, axinos are sufficiently long-lived that they almost always decay after 
neutralino freeze-out
$T_{\text{fr}}\sim m_{\tz_1}/25$~\cite{blrs}.
Axinos, produced at a sufficient rate, may induce
{\it neutralino re-annihilation} at the axino decay temperature
$T_D^{\ta}=\sqrt{\Gamma_{\ta}M_P}/(\pi^2g_*(T_D^{\ta})/90)^{1/4}$
(here, $M_P$ is the reduced Planck mass $\simeq 2\times 10^{18}$ GeV).
Since the re-annihilation Yield $Y_{\tz_1}^{re-ann}\equiv n_{\tz_1}/s $ (where $n_{\tz_1}$ is number density 
and $s$ is entropy density) is inversely proportional
to $T_D$, then when $T_D<T_{fr}$, re-annihilation also augments the neutralino abundance~\cite{ckls,blrs} 

Saxions may be produced thermally~\cite{gs}
(again proportional to $T_R/f_a^2$) or via coherent oscillations~\cite{senami}
(proportional to $f_a^2$, so important at very large $f_a\sim 10^{14}-10^{16}$ GeV). They may decay via
$s\to gg$, which leads to entropy dilution of all relics present at the time of decay, or to $\tg\tg$
or other SUSY modes, which can augment the neutralino abundance. Depending on the combinations of PQ
charges $q_i$ and PQ vevs $v_i$, saxions should also decay via $s\to aa$ or $s\to \ta\ta$
for $\xi \equiv \sum_i q_i^3v_i^2/f_a^2 \sim 1$.
The first of these leads to production of dark radiation which is stringently limited
by WMAP/Planck~\cite{wmap9} parametrized in terms of the number of additional
neutrinos present in the universe: $\Delta N_{eff}<1.6$ at 95\% CL.
If the saxion or axino decays occur much after $T_{\text{BBN}}\sim 1$ MeV, then
light elements produced during Big Bang Nucleosynthesis may become dis-associated leading also to
severe constraints~\cite{bbn}.

Alternatively to SUSY KSVZ, in the SUSY DFSZ model~\cite{dfsz,kimnilles,chun,bci} the
PQ superfield couples directly to the Higgs superfields carrying non-trivial PQ charges:
\be \label{WDFSZ}
W_{\text{DFSZ}}\ni \lambda \frac{S^2}{M_P} H_u H_d .
\ee
Here, $S$ is a Minimal Supersymmetric Standard Model (MSSM) singlet but carries
a PQ charge and contains the axion field.
An advantage of this approach is that it provides a solution to the SUSY $\mu$ problem~\cite{kimnilles}:
since the $\mu$ term is supersymmetric, one expects $\mu\sim M_P$ in contrast to
phenomenology (Eq.~(\ref{eq:mZsSig})) which requires $\mu\sim m_{\text{weak}}$.
In this Kim-Nilles solution, PQ charge assignments to the Higgs fields imply that the usual superpotential $\mu$ term
is forbidden.
Upon breaking of PQ symmetry, the field $S$ receives a vev $\langle S\rangle\sim f_a$, so that an
effective $\mu$ term is generated with $\mu\sim \lambda f_a^2/M_P\sim \lambda m_{3/2}$.
For $\lambda f_a^2/M_P\sim 100$ GeV, then one may generate $\mu\sim 100$ GeV in accord with naturalness whilst
$m_{\tq}\sim m_{3/2}\sim 10$ TeV in accord with LHC constraints and in accord with at least a partial
decoupling solution to the SUSY flavor, $CP$ and gravitino problems~\cite{dine}.

In SUSY DFSZ,  the direct coupling of the axion supermultiplet to the Higgs superfields
leads to thermal production rates which are independent of $T_R$~\cite{chun, bci}. The saxion and axino
thermal yields are then given by
\bea
Y_s^{\text{TP}} &\simeq & 10^{-7}\zeta_s\left(\frac{\mu}{\text{TeV}}\right)^2\left(\frac{10^{12}\ {\rm GeV}}{f_a}
\right)^2\\
Y_{\ta}^{\text{TP}} &\simeq & 10^{-7}\zeta_{\ta}\left(\frac{\mu}{\text{TeV}}\right)^2\left(\frac{10^{12}\ {\rm GeV}}
{f_a}\right)^2
\eea
where the $\zeta_i$ are model-dependent constants of order unity.
Saxions can also be produced via coherent oscillations with a yield given by
\be
Y_s^{\rm CO}= 1.9\times 10^{-6}\left(\frac{\rm GeV}{m_s}\right)
\left(\frac{{\rm min}[T_R,T_s]}{10^7\mbox{ GeV}}\right)
\left(\frac{f_a}{10^{12}\mbox{ GeV}}\right)^2
\ee
assuming an initial saxion field amplitude of $s_0\sim f_a$.
Along with these,
neutralinos will be produced via thermal freeze-out as usual given by
\be
Y_{\tz_1}^{\text{fr}}\simeq \frac{\left( 90/\pi^2g_*(T_{\text{fr}})\right)^{1/2}}{4\langle\sigma v\rangle M_PT_{\text{fr}}}
\label{eq:Yfr}
\ee
where $T_{\text{fr}}$ is the freeze-out temperature ($\sim m_{\tz_1}/25$) and 
$\langle\sigma v\rangle$ is the thermally averaged
neutralino annihilation cross section times relative velocity.
Axions will also be produced at the QCD phase transition via coherent oscillations~\cite{vacmis} given by
\be
\Omega_a^{\text{std}} h^2\simeq 0.23 f(\theta_i)\theta_i^2
\left(\frac{f_a/N_{\rm DW}}{10^{12}\ {\rm GeV}}\right)^{7/6}
\label{eq:Oh2axionstd}
\ee
where the misalignment angle $0< \theta_i<\pi$, $N_{\rm DW}$ is the domain wall number, and $f(\theta_i)$ is the anharmonicity
factor parametrized as $f(\theta_i)=\left[\ln\left(\frac{e}{1-\theta_i^2/\pi^2}\right)\right]^{7/6}$.

Along with the above processes, neutralinos can be produced via axino decays. In SUSY DFSZ,
the dominant modes include: $\ta\to \tz_i \phi$ (where $\phi=h,H,A$),
$\tz_i Z$ ($i=1-4$), $\tw_j^\pm H^\mp$ 
and $\tw_j^\pm W^\mp$ ($j=1-2$).
Summing over decay modes and neglecting phase space factors, the axino width is
\be
\Gamma_{\ta}\sim \frac{c_H^2}{4\pi}\left(\frac{\mu}{v_{PQ}}\right)^2m_{\ta} ,
\ee
where $c_H$ is an order one parameter for the axino (saxion) coupling arising from Eq.~(\ref{WDFSZ}) and
$v_{PQ}\equiv\sqrt{\sum_iq_i^2v_i^2}\sim f_a$ in terms of PQ charges and vevs.
This tends to greatly exceed the value obtained in SUSY KSVZ.

\begin{figure}[tbp]
\postscript{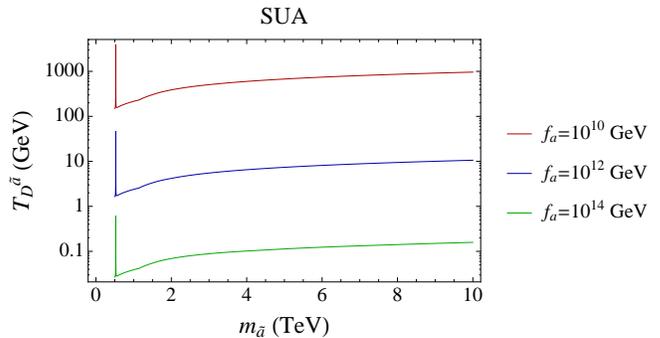}{1.0}
\caption{Decay temperature of DFSZ axinos vs.\ PQ scale $f_a$
for the SUA benchmark.
\label{fig:TDaxno}}
\end{figure}

For illustration, we adopt the Standard Underabundance SUSY benchmark (SUA) scenario
from Ref.~\cite{rns,bbl} which assumes a Higgsino mass scale $\mu\sim 150$ GeV
while sparticles which do not contribute to naturalness are at high masses $\gg$ 1 TeV, safely
beyond the LHC reach.
Our results are hardly sensitive to the selected benchmark so long as
$|\mu |\sim m_Z$ while the remaining sparticles are quite heavy.
The axino decay temperature for the SUA point using the complete decay widths
is shown in Fig.~\ref{fig:TDaxno}. 
The spike is due to a resonant axino-neutralino mixing effect when 
$|m_{\ta}-m_{\widetilde{Z}_i}|\lesssim \mu v/f_a$ and can be ignored in our analyses.
Comparing against the neutralino freeze-out temperature for a 135 GeV Higgsino-like LSP 
($T_{fr}\sim 5$ GeV), we see that $T_D^{\ta}$ tends to {\it exceed} $T_{fr}$ for $f_a\alt 10^{12}$ GeV. 
In this case, the axino-produced
neutralinos will thermalize and their abundance is determined by the usual thermal freeze-out.
For higher $f_a\agt 10^{12}$, the axinos decay after neutralino freeze-out. 
The injection of a huge population of decay-produced neutralinos into the cosmic plasma at temperatures well 
below freeze-out leads to neutralino re-annihilation at $T_D^{\ta}$.
Solving the Boltzmann equation for the neutralino re-annihilation abundance
leads to a Yield value given by~\cite{ckls,blrs}\footnote{For an explicit derivation, 
see Eq'ns 7.8-7.10 of Ref.~\cite{blrs}.} 
\be
Y_{\tz_1}^{\text{re-ann}}\simeq 
\frac{\left( 90/\pi^2g_*(T_D^{\ta})\right)^{1/2}}{4\langle\sigma v\rangle M_P T_D^{\ta}}
\label{eq:reann}
\ee
which essentially replaces $T_{\text{fr}}$ by $T_D^{\ta}$ in Eq.~(\ref{eq:Yfr}).
Since $T_D^{\ta}<T_{\text{fr}}$, this gives an
{\it increased} yield over the thermal expectation.
If the temperature at which the axino density equals the radiation density,
$T_e^{\ta}=4m_{\ta}Y_{\ta}/3$, is smaller than the decay temperature, axinos
temporarily dominate the energy density of the universe. 
Axino domination rarely happens in SUSY DFSZ since then a huge value of 
$\mu \agt 10^{-4} f_a$ is required.

\begin{figure}[tbp]
\postscript{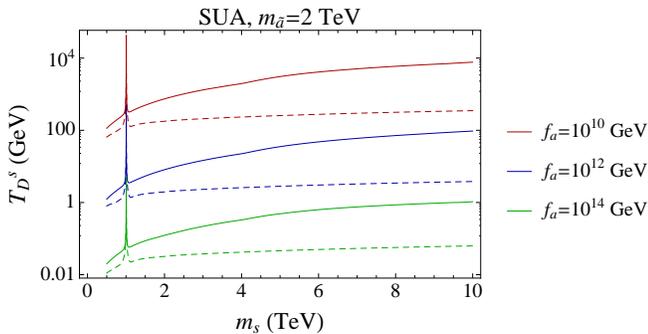}{1.0}
\caption{Decay temperature of DFSZ saxions vs.\ PQ scale $f_a$
for the SUA benchmark with $m_{\ta}=2$ TeV.
Solid curves have $\xi =1$ while dashed curves have $\xi =0$.
\label{fig:TDsaxion}}
\end{figure}

Saxions, produced thermally and non-thermally, can decay via
$s\to hh$, $HH$, $hH$, $AA$, $H^+H^-$, $ZZ$, $W^+W^-$, $ZA$, $W^\pm H^\mp$, $\tz_i\tz_{i'}$,
$\tw_j^\pm\tw_{j'}^\mp$, and also to fermions and sfermions (complete decay formulae are given in \cite{dfsz2}). 
For large $m_s$, the width is dominated by
\be
\Gamma (s\to\ {\rm Higgsinos})\simeq \frac{c_H^2}{32\pi}\left(\frac{\mu}{v_{PQ}}\right)^2m_s .
\ee
In addition, for $\xi\sim 1$, the decay $s\to aa$ may be sizable, leading to dark radiation~\cite{bbl},
or $s\to\ta\ta$ may occur as well, further augmenting the LSP abundance.
The saxion decay temperature $T_D^s$ is shown in Fig.~\ref{fig:TDsaxion}.
The spike is due to a resonant saxion-Higgs mixing effect when $|m_s^2-m_{h,H}^2|\lesssim B\mu v/f_a$ and can be ignored in our analyses.
For $f_a\alt 10^{12}$ GeV, saxions tend to decay before the neutralino freeze-out.
A comparison of the saxion radiation  equality temperature $T_e^s$ against the decay temperature $T_D^s$
shows that saxions  dominate the energy density of the universe only
when $f_a\agt 10^{14}$ GeV for which $Y_s^{\text{CO}}$ is large enough.

\begin{figure}[tbp]
\postscript{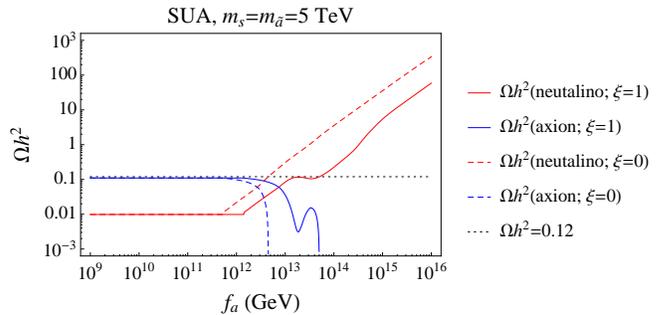}{1.0}
\caption{Neutralino and axion relic abundance from the SUSY DFSZ
axion model versus PQ scale $f_a$ for the SUA benchmark point.
\label{fig:Oh2}}
\end{figure}

\begin{figure}[tbp]
\postscript{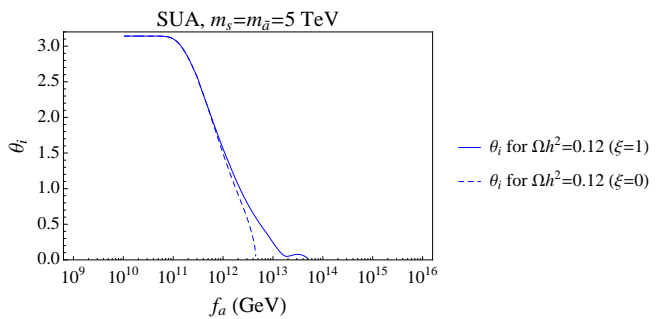}{1.0}
\caption{Axion field misalignment angle vs. $f_a$
which is required to saturate mixed axion-neutralino abundance for
$\xi=0$ (dashed) and $\xi=1$  (solid).
\label{fig:theta}}
\end{figure}

Let us now examine the contributions of neutralinos and axions to the observed dark matter density
expected in the SUSY DFSZ model. Our main result is shown in Fig.~\ref{fig:Oh2} assuming $m_{\ta}=m_s =5$ TeV.
We adopt the range of $f_a$ starting at $10^9$ GeV (stellar cooling arguments and supernovae analyses require $f_a\agt 10^9$ GeV),
and proceed to values $f_a\sim 10^{16}$ GeV (far above the naive closure bound $f_a\alt 10^{12}$ GeV 
gained from Eq.~(\ref{eq:Oh2axionstd}) using $\theta_i=1$). 
The neutralino abundance $\Omega_{\tz_1}h^2\approx 0.01$ is given by the standard thermal freeze-out 
over a {\it large range} of $f_a$ extending all the way up to $f_a\sim 10^{12}$ GeV.
In this regime, the axion abundance can always be found by adjusting $\theta_i$
such that the summed abundance meets the measured value: $\Omega_{\tz_1}h^2+\Omega_ah^2 =0.12$.
The required value of $\theta_i$ is shown in Fig.~\ref{fig:theta}. For very low $f_a\sim 10^9$ GeV, a large
value of $\theta_i\sim \pi$ is required, and $\Omega_a h^2$ is dominated by the anharmonicity term.
As $f_a$ increases, the assumed initial axion field value $\theta_i f_a$ increases, so the required
misalignment angle $\theta_i$ decreases. Values of $\theta_i\sim 1$ are found around $f_a\sim 10^{12}$ GeV
for both $\xi =0$ and 1. In this entire region with $f_a\sim 10^9-10^{12}$ GeV, we expect from natural SUSY
that the relic Higgsino abundance lies at the standard freezeout value, comprising about 5-10\% of the total
dark matter density, while axions would comprise 90-95\% of the abundance.
Thus, over the commonly considered range of $f_a$, we expect
{\it mainly axion cold dark matter from natural SUSY},
along with a non-negligible fraction of Higgsino-like WIMPs.

As $f_a$ increases beyond $10^{12}$ GeV, axinos and saxions decay later than
the neutralino freeze-out so that axino and saxion decays to SUSY particles 
add to the WIMP abundance, and $\Omega_{\tz_1}h^2$ begins to increase. 
The exact value of $\Omega_{\tz_1}h^2$ is given by the re-annihilation abundance (Eq.~(\ref{eq:reann})) at
temperatures $T_D<T_{\text{fr}}$ and so the neutralino abundance can be seen
as the rising curves of $\Omega_{\tz_1}h^2$ in Fig.~\ref{fig:Oh2} for both $\xi=0$ and $\xi=1$.
The rapid rise of the  neutralino abundance as $f_a$ increases results in excluding
the region of $f_a\agt 5\times 10^{12}$ GeV ($5\times10^{13}$ GeV) by dark matter overproduction for 
all values of $\xi=0-1$.
If we increase $m_{\ta}=m_s$ to 10 (20) TeV, then the upper bound on $f_a$ moves to $6\times 10^{12}$ 
($9\times 10^{12}$) GeV for $\xi=0$. 
For $\xi=0$, the $\Omega_{\tz_1}h^2$ curve  rises steadily with $f_a$ due to increasing production of
saxions from coherent oscillations and their dominant decays to SUSY particles. This leads to subsequent neutralino
re-annihilation at decreasing temperatures $T_D^s$. For $\xi =1$, the dominant saxion decay mode is $s\to aa$,
and decay-produced neutralinos come mainly from thermal axino production which decreases as
$f_a$ increases. One sees that $\Omega_{\tz_1}h^2$ turns over and briefly reaches $\Omega_{\tz_1}h^2\simeq 0.1$
at $f_a\sim 3\times 10^{13}$ GeV before beginning again a rise due to increasing non-thermal saxion
production. It is important to note that for  $\xi \sim 1$ and $f_a\agt 10^{14}$ GeV,  too much dark radiation
is produced ($\Delta N_{eff}>1.6$) (not shown here) and thus large $f_a$ is excluded by overproduction of both dark radiation
and WIMPs.

{\it Summary:} Supersymmetry with not too heavy top squarks, low Higgsino mass $\mu\sim 100-200$ GeV
and PQWW solution to the strong CP problem successfully avoids high fine-tuning in both the electroweak and 
QCD sectors of the theory while evading LHC constraints. 
The SUSY DFSZ model, wherein Higgs superfields carry PQ charge, also provides a solution to
the SUSY $\mu$ problem. In such models, over a large range of PQ breaking scale $f_a\sim 10^9-10^{12}$ GeV,
saxions and axinos typically decay before neutralino freeze-out so that the Higgsino portion of dark matter is
expected to lie in the 5-10\% range while axions would comprise the remainder: 90-95\%. The relic Higgsinos
ought to be detectable at ton scale noble liquid detectors, even with a depleted local abundance, while
indirect detection should be more limited since expected rates go as the depleted abundance squared~\cite{bbm}.
Prospects are bright for microwave cavity detection of axions since the range of $f_a$ where mainly axion
dark matter is expected should be accessible to experimental searches~\cite{axsearch}.
While corroborative searches for natural SUSY with light Higgsinos is limited at the LHC~\cite{wp},
a definitive Higgsino search should be possible at
$e^+e^-$ colliders with $\sqrt{s}$ up to $500-600$ GeV.

{\it Acknowledgements:}
We thank V. Barger for comments on the manuscript and A. Lessa for collaboration on the early
phase of this work.
This work was supported in part by the US Department of Energy, Office
of High Energy Physics.


%
\end{document}